\documentclass[aps,12pt,final,notitlepage,oneside,onecolumn,nobibnotes,nofootinbib,superscriptaddress,noshowpacs,centertags]{revtex4}
\usepackage{graphics}
\begin{document}
\title{Correlator of Heavy--Light Quark Currents in HQET\\in the Large $\beta_0$ Limit}
\author{\copyright{} 2026\hspace{8mm}\firstname{A.~G.}~\surname{Grozin}}%
\email{A.G.Grozin@inp.nsk.su}%
\affiliation{Budker Institute of Nuclear Physics, Novosibirsk, Russia}%
\affiliation{Bogoliubov Laboratory of Theoretical Physics, Joint Institute for Nuclear Research, Dubna, Russia}
\begin{abstract}
\textbf{Abstract --}
The perturbative contribution to the correlator of two heavy-light quark currents in HQET
expanded in light-quark masses up to quadratic terms is calculated at the leading order in $1/\beta_0$.
Ultraviolet and infrared renormalon poles of Borel images of the Wilson coefficients are discussed.
\end{abstract}
\maketitle
\section{Introduction}
QCD problems with a single heavy quark (having the pole mass $M$)
can be handled by Heavy Quark Effective Theory
(HQET, see, e.g., \cite{Manohar:2000dt,Grozin:2004yc}).
At the leading order in $1/M$ the heavy quark spin does not interact with the gluon field
(chromomagnetic interaction is ${\sim}1/M$)
so that the heavy quark spin can be switched off.
The HQET Lagrangian with the fields $h$ (spin $\frac{1}{2}$) and $\varphi$ (spin 0)
in the static-quark rest frame,
\begin{equation}
L = h^+ i D_0 h + \varphi^* i D_0 \varphi\,,
\label{L}
\end{equation}
has superflavor symmetry $SU(3)$~\cite{Georgi:1990ak}.
The coordinate-space propagator of the spinless heavy quark is
\begin{equation}
S_0(x) = \delta(\mathbf{x}) S_0(x^0)\,,\quad
S_0(t) = - i \theta(t)\,,
\label{Scoord}
\end{equation}
i.e., the quark stays where it has been created;
it propagates only forward in time,
so that its line cannot form loops.
The momentum-space propagator
\begin{equation}
S_0(p) = \frac{1}{p^0 + i0}
\label{Smom}
\end{equation}
depends only on $p^0$, but not on $\mathbf{p}$.

We consider heavy-light quark currents
\begin{equation}
j_{P0} = \frac{1 + P\gamma^0}{2} \varphi^*_0 q_0\,,\quad
P = \pm1\,,
\label{j0}
\end{equation}
where $\varphi^*$ is the spinless heavy antiquark field,
$q$ is the field of a light quark,
$P$ is the current parity,
the index 0 means unrenormalized quantities.
The $P = +1$ current has quantum numbers of $B$, $B^*$ mesons
(they are identical at the leading order in $1/M$),
and the $P = -1$ one --- of the $P$-wave $0^+$, $1^+$ mesons.
The correlators are defined as
\begin{equation}
\left\langle T j_{P0}(x) \bar{\jmath}_{P0}(0) \right\rangle
= \delta(\mathbf{x}) P \frac{1 + P\gamma^0}{2} \Pi_{P0}(x^0)\,.
\label{Pi0}
\end{equation}
Analytically continuing $\Pi_{P0}(t)$ from the $t>0$ half-axis to the $t = -i\tau$, $\tau>0$ half-axis
we obtain the Euclidean correlators $\Pi_{P0}(\tau)$.

The operator product expansion (OPE) for correlators of the renormalized currents
in momentum and coordinate spaces is
\begin{eqnarray}
\Pi_P(\omega,\mu) &=& \sum_{\mathcal{O}} C_{P,\mathcal{O}}(\omega,\mu) \langle \mathcal{O}(\mu)\rangle\quad
\bigl(-\omega \gg \Lambda_{\overline{\text{MS}}}\bigr)\,,
\label{OPE1}\\
\Pi_P(\tau,\mu) &=& \sum_{\mathcal{O}} C_{P,\mathcal{O}}(\tau,\mu) \langle \mathcal{O}(\mu) \rangle\quad
\biggl(\tau\ll\frac{1}{\Lambda_{\overline{\text{MS}}}}\biggr)\,.
\label{OPE2}
\end{eqnarray}
The coefficient functions $C_{\mathcal{O}}$ contain hard momenta ${>}\mu$ (short distances ${<}1/\mu$);
the matrix elements $\langle\mathcal{O}\rangle$ contain soft momenta ${<}\mu$ (long distances ${>}1/\mu$).
In schemes based on dimensional regularization the wall separating hard and soft momenta is fuzzy.
Coefficient functions contain, in addition to the main hard contributions,
also soft contributions leading to infrared (IR) renormalons and the corresponding ambiguities;
matrix elements contain, in addition to the main soft contributions,
also hard contributions leading to ultraviolet (UV) renormalons and the corresponding ambiguities.
They are artifacts of scale separation --- they are absent in the full correlator.
If one changes the prescription for summing divergent perturbative series for a coefficient function
(thus changing the value of the sum),
one has to change the values of higher-dimensional vacuum condensates accordingly.

We use $\overline{\text{MS}}$ renormalization scheme, $\mu$ is the normalization scale;
\begin{equation}
\gamma_a = \frac{d\log Z_a}{d\log\mu} = \sum_{l=1}^\infty \gamma_{a,l-1} \biggl(\frac{\alpha_s}{4\pi}\biggr)^l\,,\quad
a = j, m;\quad
\beta = \frac{1}{2} \frac{d\log Z_\alpha}{d\log\mu} = \sum_{l=1}^\infty \beta_{l-1} \biggl(\frac{\alpha_s}{4\pi}\biggr)^l\,.
\label{gammabeta}
\end{equation}
The anomalous dimension of $C_{P,m^n}$ ($n \in [0,2]$) is
\begin{equation}
\gamma_n = 2 \gamma_j - n \gamma_m\,.
\label{gamman}
\end{equation}

The correlator $\Pi_P(\omega,\mu)$ of the renormalized currents $j_P(\mu)$ contains UV divergences $1/\varepsilon^n$;
their coefficients are polynomial in $\omega$.
Therefore, the coefficients of the divergences in $\Pi_P(t,\mu)$ contain $\delta^{(n)}(t)$;
they disappear in the analytical continuation to $\tau$.
Divergences are also absent in the spectral density
\begin{equation}
\rho_P(\omega,\mu) = \frac{\Pi_P(\omega+i0,\mu) - \Pi_P(\omega-i0,\mu)}{2\pi}
\label{rho}
\end{equation}
--- polynomials in $\omega$ don't contribute to the discontinuity at the cut $\omega>0$.

Here we consider operators with dimensions $D \le 2$:
\begin{equation}
\mathcal{O}_0 = 1\,,\quad
m_0\,,\quad
m_0^2\,,\quad
\sum m_{i0}^2\,,
\label{D2}
\end{equation}
where $m$ is the mass of the light quark $q$ in the current $j$~(\ref{j0}),
and $m_i$ are the masses of all light flavors.
In other words, we consider the perturbative contribution to the correlators.
At dimension $D=3$ the quark condensate $\bar{q}q$ mixes with $m^3$,
and the analysis becomes more complicated.
The coefficients $C_{\mathcal{O}}$ with even dimensionalities $D$ don't contain $P$;
for odd $D$ $C_{\mathcal{O}} \propto P$.
In what follows we set $P = +1$;
for $P = -1$ it is sufficient to revert the sign of $C_m$.
The perturbative contribution to the correlator has been calculated
at 2~\cite{Broadhurst:1991fc,Bagan:1991sg,Neubert:1991sp},
3~\cite{Chetyrkin:2021qvd} and 4~\cite{Grozin:2024dhk} loops.

\section{Large $\beta_0$ limit}
Coefficients in the perturbative series for $C_{m^n,0}(\tau)$ ($n \in [0,2]$) are polynomials in $T_F n_f$:
\begin{equation}
A_{n0}(\tau) = \frac{C_{m^n,0}(\tau)}{C_{m^n,0}^{(1)}(\tau)}
= 1 + \sum_{l=1}^\infty \sum_{k=0}^{l-1} a'_{nlk} (T_F n_f)^k \biggl(\frac{g_0^2}{(4\pi)^{d/2}}\biggr)^{\!\!l}
\label{nf}
\end{equation}
(here $C_{m^n,0}^{(1)}(\tau)$ is the 1-loop contribution, it is convergent).
We can re-write%
\footnote{$n_f$ can also appear, e.g., via $n_f d_F^{abcd}$, so that it is multiplied by a quartic Casimir.
Then it cannot be expressed via $\beta_0$.
Such terms first appears at the order $g_0^8$, i.e. $1/\beta_0^4$;
they are not relevant for us here.}
these polynomials as polynomials in $\beta_0$:
\begin{equation}
A_{n0}(\tau) = 1 + \sum_{l=1}^\infty \sum_{k=0}^{l-1} a_{nlk} \beta_0^k \biggl(\frac{g_0^2}{(4\pi)^{d/2}}\biggr)^{\!\!l}\,.
\label{beta0}
\end{equation}
Here we consider the large $\beta_0$ limit:
$\beta_0 \alpha_s \sim 1$, $1/\beta_0$ is our small parameter
(see~\cite{Beneke:1998ui} and Chapter~8 in~\cite{Grozin:2004yc}).
We consider the first order in $1/\beta_0$
(in some problems is appears possible to calculate $1/\beta_0^2$ corrections,
but only in problems containing some factor which simplifies the analysis considerably).

The unrenormalized quantity $A_{n0}(\tau)$ can be written in the form~\cite{Palanques-Mestre:1983ogz,Broadhurst:1992si}:
\begin{equation}
A_{n0}(\tau) = 1 + \frac{C_F}{\beta_0} \sum_{l=1}^\infty \frac{F_n(\varepsilon,l\varepsilon)}{l}
\biggl[\frac{\beta_0 g_0^2}{(4\pi)^{d/2}} \biggl(\frac{\tau e^{\gamma}}{2}\biggr)^{\!\!2\varepsilon}
e^{-\gamma\varepsilon} \frac{D(\varepsilon)}{\varepsilon}\biggr]^l
+ \mathcal{O}\biggl(\frac{1}{\beta_0^2}\biggr)\,.
\label{A0}
\end{equation}
Here
\begin{equation}
D(\varepsilon) = 6 e^{\gamma\varepsilon} \frac{\Gamma(1+\varepsilon) \Gamma^2(2-\varepsilon)}{\Gamma(4-2\varepsilon)}
= 1 + \frac{5}{3} \varepsilon + \cdots,
\label{De}
\end{equation}
$\gamma$ is the Euler constant.
Re-expressing this result via the renormalized coupling constant
\begin{equation}
b = \beta_0 \frac{\alpha_s(\mu)}{4\pi} \sim 1\,,
\label{b}
\end{equation}
we have
\begin{equation}
A_{n0}(\tau) = 1 + \frac{C_F}{\beta_0} \sum_{l=1}^\infty \frac{F_n(\varepsilon,l\varepsilon)}{l}
\biggl[\frac{b}{\varepsilon+b} \biggl(\frac{\mu\tau e^{\gamma}}{2}\biggr)^{\!\!2\varepsilon} D(\varepsilon)\biggr]^l
+ \mathcal{O}\biggl(\frac{1}{\beta_0^2}\biggr)\,.
\label{Ab}
\end{equation}
It is convenient to set $\mu = \mu_\tau$:
\begin{equation}
\mu_\tau = \frac{2 e^{-\gamma}}{\tau} D(\varepsilon)^{-1/(2\varepsilon)} \to \frac{2}{\tau} e^{-\gamma-5/6}\,,
\label{mut}
\end{equation}
then
\begin{equation}
A_{n0}(\tau) = 1 + \frac{C_F}{\beta_0} \sum_{l=1}^\infty \frac{F_n(\varepsilon,l\varepsilon)}{l}
\biggl(\frac{b}{\varepsilon+b}\biggr)^{\!\!l} + \mathcal{O}\biggl(\frac{1}{\beta_0^2}\biggr)\,.
\label{Amut}
\end{equation}
The functions $F_n(\varepsilon,u)$ contain all information about the $1/\beta_0$ terms; they are regular at the origin:
\begin{equation}
F_n(\varepsilon,u) = \sum_{i=0}^\infty \sum_{j=0}^\infty F_{n,ij} \varepsilon^i u^j\,.
\label{Feu}
\end{equation}
Expanding also $(b/(\varepsilon+b))^l$ in $b$ we obtain quadruple sums for $A_{n0}(\tau)$.

Let's select $\varepsilon^{-1}$ terms from these sums.
All coefficients $F_{n,ij}$ cancel except $F_{n,i0}$~\cite{Palanques-Mestre:1983ogz};
these terms, i.e. the functions $F_n(\varepsilon,0)$, give us the anomalous dimensions $\gamma_n$:
\begin{equation}
\gamma_n(b) = - 2 C_F \frac{b}{\beta_0} F_n(-b,0) + \mathcal{O}\biggl(\frac{1}{\beta_0^2}\biggr)\,.
\label{gamma}
\end{equation}
We have reproduced the result~\cite{Broadhurst:1994se}
\begin{equation}
\gamma_j(b) = - \frac{1}{2} \gamma_m(b) + \mathcal{O}\biggl(\frac{1}{\beta_0^2}\biggr)
= - C_F \frac{b}{\beta_0} \frac{1+\frac{2}{3}b}{B(2+b,2+b) \Gamma(3+b) \Gamma(1-b)} + \mathcal{O}\biggl(\frac{1}{\beta_0^2}\biggr)\,,
\label{gammaj}
\end{equation}
so that $\gamma_n = 2 (n+1) \gamma_j + \mathcal{O}(1/\beta_0^2)$.
In particular,
\begin{equation}
\gamma_{n0} = - 2 C_F F_n(0,0)\,,\quad
F_n(0,0) = 3 (n+1)\,.
\label{gamma0}
\end{equation}
Terms of $A_{n0}(\tau)$ with $1/\varepsilon^{-n}$, $n>1$ contain no new information ---
they are unambiguously fixed by the $\varepsilon^{-1}$ terms
and the requirement that anomalous dimensions are finite at $\varepsilon\to0$;
hence they are also determined by $F_{n,i0}$.

The solution of the renormalization group (RG) equation for the renormalized correlators $A_n(\tau,\mu)$ ($n \in [0,2]$)
can be written as
\begin{equation}
A_n(\tau,\mu) = \hat{A}_n(\tau) \biggl(\frac{\alpha_s(\mu)}{\alpha_s(\mu_\tau)}\biggr)^{\!\!\gamma_{n0}/(2\beta_0)} K_n(\alpha_s(\mu))\,,\quad
K_n(\alpha_s) = \exp \int_0^{\alpha_s} \frac{d\alpha_s}{\alpha_s}
\biggl(\frac{\gamma_n(\alpha_s)}{2\beta(\alpha_s)} - \frac{\gamma_{n0}}{2\beta_0}\biggr)\,,
\label{RG}
\end{equation}
where $\hat{A}_n(\tau)$ is an RG invariant ($\mu$ independent).
Let's select $\varepsilon^0$ terms from the quadruple sums for $A_{n0}(\tau)$.
All coefficients $F_{n,ij}$ cancel except $F_{n,i0}$ and $F_{n,0j}$~\cite{Broadhurst:1992si}.
The former ones (i.e. $F_n(-b,0)$) produce $K_n(b)$,
and the latter ones (i.e. $F_n(0,u)$) produce $\hat{A}_n(\tau)$:
\begin{equation}
\hat{A}_n(\tau) = 1 + \frac{C_F}{\beta_0} \int_0^\infty du\,e^{-u/b} S_n(u) + \mathcal{O}\biggl(\frac{1}{\beta_0^2}\biggr)\,,\quad
S_n(u) = \frac{F_n(0,u)-F_n(0,0)}{u}
\label{hatA}
\end{equation}
(here $b = b(\mu_\tau)$).

In order to find the functions $F_n(\varepsilon,u)$ one needs to calculate the 2-loop diagrams
\begin{equation}
\raisebox{-1mm}{\includegraphics{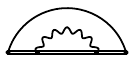}}\,,\quad
\raisebox{-1mm}{\includegraphics{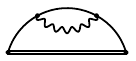}}\,,\quad
\raisebox{-1mm}{\includegraphics{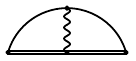}}\,,
\label{dia}
\end{equation}
where the denominator of the gluon propagator is raised to the power $1+u-\varepsilon$.
They are expressible via $\Gamma$ functions and a single non-trivial integral
\begin{equation}
I(x) = \raisebox{1mm}{\begin{picture}(22,9.5)
\put(28,4.75){\makebox(0,0){\includegraphics{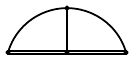}}}
\put(28.5,4.75){\makebox(0,0)[l]{$x$}}
\end{picture}}\hspace{13mm},
\label{BB}
\end{equation}
which can be expressed~\cite{Beneke:1994sw} via a hypergeometric function ${}_3F_2$ of the unit argument.

The functions $S_n(u)$ have poles at $u>0$, so that the integrals~(\ref{hatA}) are ill-defined.
Let's say there is a renormalon pole at $u_0 > 0$:
\begin{equation}
S_n(u) \sim \frac{r}{u_0 - u}\,.
\label{pole}
\end{equation}
Then the renormalon ambiguity of $\hat{A}_n(\tau)$~(\ref{hatA}), and hence of $A_n(\tau,\mu)$~(\ref{RG}),
can be estimated by the residue of the integrand:
\begin{equation}
\Delta A_n(\tau,\mu) = \frac{C_F}{\beta_0} r \biggl(\frac{\Lambda_{\overline{\text{MS}}}}{\mu_\tau}\biggr)^{\!\!2 u_0}
= \frac{C_F}{\beta_0} r \biggl(e^{\gamma+5/6} \Lambda_{\overline{\text{MS}}} \frac{\tau}{2}\biggr)^{\!\!2 u_0}\,.
\label{Delta}
\end{equation}
For example, one can choose the principal-value prescription,
i.e.\ cut out the interval $[u_0-\delta,u_0+\delta]$, $\delta\to0$.
But if one changes the prescription, e.g., cuts the interval $[u_0-\delta,u_0+2\delta]$,
then the value of the integral changes by a quantity of the order of the residue.
The renormalon contribution~(\ref{pole}) expanded in $u$ and substituted into~(\ref{hatA})
produces the contribution to the perturbative series
\begin{equation}
\hat{A}_n(\tau) = 1 + \frac{C_F}{\beta_0} \sum_{l=1}^\infty a_l b^l + \mathcal{O}\biggl(\frac{1}{\beta_0^2}\biggr)\,,\quad
a_l \sim r \frac{(l-1)!}{u_0^l}
\label{asym}
\end{equation}
with the coefficients growing factorially.
If $b \ll 1$ then the series terms first decrease, then reach the minimum and start to grow.
It seems reasonable to sum such an asymptotic series up to the minimum term,
and to declare this term to be the ambiguity of the result.
The series terms $a_l b^l \sim r (l b/(e u_0))^l$ reach the minimum at $l \sim u_0/b$;
it is equal to~(\ref{Delta}).

\section{Euclidean--time correlator}
The functions $S_n(u)$ ($n \in [0,2]$) have UV renormalon poles at $u=\frac{1}{2}$:
\begin{equation}
S_n(u) \sim e^{-\gamma} \frac{4}{\frac{1}{2} - u}\,.
\label{UVren}
\end{equation}
This gives the renormalon ambiguities of $A_n(\tau,\mu)$, and hence of the correlator $\Pi(\tau,\mu)$:
\begin{equation}
\frac{\Delta\Pi(\tau,\mu)}{\Pi(\tau,\mu)} = 2 \frac{C_F}{\beta_0} e^{5/6} \Lambda_{\overline{\text{MS}}} \tau
= - \Delta\bar{\Lambda}\,\tau\,,\quad
\Delta\bar{\Lambda} = - 2 \frac{C_F}{\beta_0} e^{5/6} \Lambda_{\overline{\text{MS}}}
\label{UVren0}
\end{equation}
is the UV renormalon ambiguity of $\bar{\Lambda}$,
the energy of the ground-state meson ($B$, $B^*$) in HQET~\cite{Beneke:1994sw}.
HQET energies are measured with respect to the heavy quark pole mass $M$.
This zero level is fuzzy due to the IR renormalon $u=\frac{1}{2}$ in $M$:
$\Delta\bar{\Lambda} = - \Delta M$.
One can choose some perfectly definite zero energy level,
but then the HQET Lagrangian will contain the residual mass term~\cite{Falk:1992fm},
and it will have an IR renormalon ambiguity.
The correlator can be written as
\begin{equation}
\Pi(\tau,\mu) = \sum_i c_i e^{- \bar{\Lambda}_i \tau},
\label{disp}
\end{equation}
where $\bar{\Lambda}_i$ are the energies of all intermediate states in the correlator;
all $\Delta\bar{\Lambda}_i = \Delta\bar{\Lambda}$, and this leads to~(\ref{UVren0}).
I.e. the UV renormalon ambiguity of the correlator~(\ref{UVren0})
is compensated by the IR renormalon ambiguity of the pole mass  $M$~\cite{Beneke:1994sw}.
If one changes the summing prescription for the coefficient functions,
one has to change $M$ accordingly;
the zero energy level will shift, thus shifting all $\bar{\Lambda}_i$.

IR renormalon poles of $S_0(u)$ are situated at integer $u\ge3$:
\begin{equation}
S_0(u) \sim - e^{-6 \gamma} \biggl(\frac{1}{36} \frac{1}{(3-u)^2} + \frac{2}{27} \frac{1}{3-u}\biggl)\,,\quad
\frac{\Delta\Pi(\tau,\mu)}{\Pi(\tau,\mu)} = - \frac{1}{864} \frac{C_F}{\beta_0} \bigl(e^{5/6} \Lambda_{\overline{\text{MS}}} \tau\bigr)^6\,.
\label{IR0}
\end{equation}
This ambiguity is compensated by the UV renormalon ambiguity of vacuum condensates.
The condensates $\langle\bar{q}q\rangle$ and $\langle\bar{q}G\sigma q\rangle$ have an opposite chirality,
their contributions are ${\propto}P$.
The gluon condensate $\langle G^2\rangle$ does not contribute to the correlator at 1 loop%
\footnote{It is easy to understand it in coordinate space in the fixed-point gauge:
the HQET propagator does not interact with gluons;
the $G^2$ correction to the massless-quark propagator $S(x,0)$ vanishes when averaged over the vacuum~\cite{Shifman:1982zt}.}.
The first condensates which can contribute are the $D=6$ condensates $\langle(\bar{q}q)^2\rangle$,
therefore the first IR renormalon is at $u=3$~\cite{Beneke:1994sw}.
The functions $S_{1,2}(u)$ have IR renormalon poles at integer $u\ge1$:
\begin{eqnarray}
S_1(u) \sim e^{-2 \gamma} \frac{6}{1-u}
- e^{-4 \gamma} \biggl(\frac{3}{2} \frac{1}{(2-u)^2} + \frac{9}{2} \frac{1}{2-u}\biggr) + \cdots,
\nonumber\\
S_2(u) \sim e^{-2 \gamma} \frac{3}{1-u}
- e^{-4 \gamma} \biggl(\frac{1}{2} \frac{1}{(2-u)^2} + \frac{9}{4} \frac{1}{2-u}\biggr) + \cdots.
\label{IRS12}
\end{eqnarray}

An explicit example of such cancellation was discussed, e.g., in~\cite{Grozin:2004ez},
where the correlator of heavy-heavy quark currents at small $q^2$ at the order $1/\beta_0$ has been considered.
The coefficient function of the unit operator contains an IR renormalon at $u=2$.
The gluon condensate can be represented as the sum of the purely soft contribution
(the gluon momentum ${<}\lambda$) which is not calculable perturbatively
and the hard tail (the momentum ${>}\lambda$) which can be calculated at the order $1/\beta_0$.
The last contribution contains an UV renormalon pole at $u=2$;
its residue multiplied by the coefficient function of the gluon condensate
cancels the IR renormalon pole of the coefficient function of the unit operator.
Such an analysis of UV renormalons of dimension 6 4-quark condensates has not been done,
therefore, we cannot check the cancellation of residues at $u=3$ in our problem explicitly.

The unrenormalized coefficient function $C_{\sum m_i^2,0}(\tau)$ can be written similarly to~(\ref{A0}):
\begin{eqnarray}
&&C_{\sum m_i^2,0}(\tau) = - \frac{16}{3} \frac{N_c C_F T_F}{(4\pi)^{d/2}} \biggl(\frac{2}{\tau}\biggr)^{\!\!1-2\varepsilon}
\label{sm2}\\
&&{}\times\biggl\{
\frac{1}{\beta_0^2} \sum_{l=1}^\infty \frac{F_\Sigma(\varepsilon,l\varepsilon)}{l}
\biggl[\frac{\beta_0 g_0^2}{(4\pi)^{d/2}} \biggl(\frac{\tau e^{\gamma}}{2}\biggr)^{\!\!2\varepsilon}
e^{-\gamma\varepsilon} \frac{D(\varepsilon)}{\varepsilon}\biggr]^l
+ \mathcal{O}\biggl(\frac{1}{\beta_0^3}\biggr)\biggr\}\,.
\nonumber
\end{eqnarray}
Re-expressing it via $b$ at $\mu = \mu_\tau$~(\ref{mut}) we have, similarly to ~(\ref{Amut}),
\begin{equation}
C_{\sum m_i^2,0}(\tau) = - \frac{16}{3} \frac{N_c C_F T_F}{(4\pi)^{d/2}} \biggl(\frac{2}{\tau}\biggr)^{\!\!1-2\varepsilon} \biggl[
\frac{1}{\beta_0^2} \sum_{l=1}^\infty \frac{F_\Sigma(\varepsilon,l\varepsilon)}{l}
\biggl(\frac{b}{\varepsilon+b}\biggr)^{\!\!l}
+ \mathcal{O}\biggl(\frac{1}{\beta_0^3}\biggr)\biggr]\,.
\label{sm2b}
\end{equation}
The function $F_\Sigma(\varepsilon,u)$ has the properties $F_\Sigma(\varepsilon,\varepsilon) = F_\Sigma(\varepsilon,0) = 0$.
The former one guarantees that there is no $l=1$ term in the sum~(\ref{sm2b}) --- such a diagram does not exist.
The latter one implies that there are no $\varepsilon^{-n}$, $n\ge1$ terms in the sum.
I.e., renormalization of this coefficient function is not needed ---
there are no terms which, after multiplying be a renormalization constant, would produce an element of our series.
The renormalized result is equal to the unrenormalized one and does not depend on $\mu$;
at $\varepsilon=0$ we have
\begin{equation}
C_{\Sigma m_i^2}(\tau) = - \frac{2 N_c C_F T_F}{3 \pi^2 \tau} \biggl[
\frac{1}{\beta_0^2} \int_0^\infty du\,e^{-u/b} S_\Sigma(u)
+ \mathcal{O}\biggl(\frac{1}{\beta_0^3}\biggr)\biggr]\,,\quad
S_\Sigma(u) = \frac{F_\Sigma(0,u)}{u}\,.
\label{sm2r}
\end{equation}
The function $S_\Sigma(u)$ has no UV renormalon pole at $u=\frac{1}{2}$;
IR renormalon poles are situated at integer $u\ge2$:
\begin{equation}
S_\Sigma(u) \sim \frac{1}{12} \frac{1}{(2-u)^2} + \frac{13}{72} \frac{1}{2-u}\,.
\label{IRSSigma}
\end{equation}

\section{Spectral density}
Taking the discontinuity of~(\ref{OPE1}) at the cut we get the spectral density
\begin{equation}
\rho(\omega,\mu) = \sum_{\mathcal{O}} R_{\mathcal{O}}(\omega,\mu) \langle \mathcal{O}(\mu) \rangle\,.
\label{OPErho}
\end{equation}
Similarly to~(\ref{A0}) we have
\begin{equation}
\tilde{A}_{n0}(\omega) = \frac{R_{m^n,0}(\omega)}{R_{m^n,0}^{(1)}(\omega)}
= 1 + \frac{C_F}{\beta_0} \sum_{l=1}^\infty \frac{\tilde{F}_n(\varepsilon,l\varepsilon)}{l}
\biggl[\frac{b}{\varepsilon+b} \biggl(\frac{\mu}{2\omega}\biggr)^{\!\!2\varepsilon} D(\varepsilon)\biggr]^l
+ \mathcal{O}\biggl(\frac{1}{\beta_0^2}\biggr)
\label{Ar0}
\end{equation}
(where $R_{m^n,0}^{(1)}(\omega)$ is the 1-loop term finite at $\varepsilon\to0$),
\begin{equation}
\tilde{F}_n(\varepsilon,u) = \frac{\Gamma(3-n-2\varepsilon)}{\Gamma(3-n-2u-2\varepsilon)} e^{2 \gamma u} F_n(\varepsilon,u)\,.
\label{Ft}
\end{equation}
It is convenient to set $\mu = \mu_\omega$:
\begin{equation}
\mu_\omega = 2\omega D(\varepsilon)^{-1/(2\varepsilon)} \to 2\omega e^{-5/6}\,,
\label{muw}
\end{equation}
then
\begin{equation}
\tilde{A}_n(\omega) = 1 + \frac{C_F}{\beta_0} \sum_{l=1}^\infty \frac{\tilde{F}_n(\varepsilon,l\varepsilon)}{l}
\biggl(\frac{b}{\varepsilon+b}\biggr)^{\!\!l}
+ \mathcal{O}\biggl(\frac{1}{\beta_0^2}\biggr)\,.
\label{Amuw}
\end{equation}
Naturally, $\tilde{F}_n(-b,0) = F_n(-b,0)$ gives~(\ref{gamma}) the same anomalous dimension $\gamma_n$~(\ref{gamman});
similarly to~(\ref{RG}) and~(\ref{hatA}),
\begin{eqnarray}
\tilde{A}_n(\omega,\mu) &=& \hat{\tilde{A}}_n(\omega) \biggl(\frac{\alpha_s(\mu)}{\alpha_s(\mu_\omega)}\biggr)^{\!\!\gamma_{n0}/(2\beta_0)}
K_n(\alpha_s(\mu))\,,
\label{RG2}\\
\hat{\tilde{A}}_n(\omega) &=&
1 + \frac{C_F}{\beta_0} \int_0^\infty du\,e^{-u/b} \tilde{S}_n(u) + \mathcal{O}\biggl(\frac{1}{\beta_0^2}\biggr)\,,\quad
\tilde{S}_n(u) = \frac{\tilde{F}_n(0,u)-\tilde{F}_n(0,0)}{u}
\label{hatAw}
\end{eqnarray}
(here $b = b(\mu_\omega)$).
If the function $\tilde{S}_n(u)$ has a pole ${\sim}r/(u_0-u)$, $u_0>0$,
this leads to the ambiguity
\begin{equation}
\Delta \tilde{A}_n(\omega,\mu) = \frac{C_F}{\beta_0} r \biggl(\frac{\Lambda_{\overline{\text{MS}}}}{\mu_\omega}\biggr)^{\!\!2 u_0}
= \frac{C_F}{\beta_0} r \biggl(e^{5/6} \frac{\Lambda_{\overline{\text{MS}}}}{2\omega}\biggr)^{\!\!2 u_0}\,.
\label{Delta2}
\end{equation}

The functions $\tilde{S}_n(u)$ ($n \in [0,2]$) have UV renormalon poles at $u=\frac{1}{2}$:
\begin{equation}
\tilde{S}_n(u) \sim \frac{4 (2-n)}{\frac{1}{2} - u}\,,\quad
\frac{\Delta R_{m^n}(\omega,\mu)}{R_{m^n}(\omega,\mu)} = (2-n) \frac{\Delta\bar{\Lambda}}{\omega}\,.
\label{UVrho}
\end{equation}
This is not surprising: $R_{m^n}(\omega) \propto \omega^{2-n}$, $\Delta\omega = \Delta\bar{\Lambda}$.
The first IR renormalon pole of $\tilde{S}_0(u)$ is at $u=3$:
\begin{equation}
\tilde{S}_0(u)\sim \frac{2}{3} \frac{1}{3-u}\,,\quad
\frac{\Delta\rho(\omega,\mu)}{\rho(\omega,\mu)} = \frac{1}{96} \frac{C_F}{\beta_0}
\biggl(e^{5/6} \frac{\Lambda_{\overline{\text{MS}}}}{\omega}\biggr)^{\!\!6}\,.
\label{IRrho}
\end{equation}
This ambiguity is compensated by the UV renormalon ambiguity
of the condensates $\langle(\bar{q}q)^2\rangle$ of dimension 6 in OPE.
The functions $\tilde{S}_{1,2}(u)$ have IR renormalon poles at $u = 2$, 3\ldots%
\footnote{The factor in~(\ref{Ft}) eliminates simple poles in~(\ref{IRS12}) and converts double poles to simple ones.}:
\begin{eqnarray}
&&\tilde{S}_1(u) \sim - \frac{6}{2-u}\,,\quad
\frac{\Delta R_m(\omega,\mu)}{R_m(\omega,\mu)} = - \frac{3}{2} \frac{C_F}{\beta_0}
\biggl(e^{5/6} \frac{\Lambda_{\overline{\text{MS}}}}{\omega}\biggr)^{\!\!4}\,,
\label{IR1}\\
&&\tilde{S}_2(u) \sim \frac{6}{2-u}\,,\quad
\frac{\Delta R_{m^2}(\omega,\mu)}{R_{m^2}(\omega,\mu)} = - 3 \frac{C_F}{\beta_0}
\biggl(e^{5/6} \frac{\Lambda_{\overline{\text{MS}}}}{\omega}\biggr)^{\!\!4}\,.
\label{IR2}
\end{eqnarray}
The IR renormalon ambiguity of $m R_m(\omega,\mu)$~(\ref{IR1}) is compensated by the UV renormalon ambiguity
of the condensate $\langle\bar{q}G\sigma q\rangle$ which has the same chirality (the contribution ${\propto}P$).
The IR renormalon ambiguity of $m^2 R_{m^2}(\omega,\mu)$~(\ref{IR2}) is compensated by the UV renormalon ambiguity
of the gluon condensate in the $m^2 \langle G^2\rangle$ contribution,
as well as of the condensates $\langle(\bar{q}q)^2\rangle$.

For $\sum m_i^2$ we have, similarly to ~(\ref{Ar0}),
\begin{eqnarray}
&&R_{\sum m_i^2,0}(\omega) = - \frac{16}{3} \frac{N_c C_F T_F}{(4\pi)^{d/2}} (2\omega)^{-2\varepsilon} \biggl\{
\frac{1}{\beta_0^2} \sum_{l=1}^\infty \frac{\tilde{F}_\Sigma(\varepsilon,l\varepsilon)}{l}
\biggl[\frac{b}{\varepsilon+b} \biggl(\frac{\mu}{2\omega}\biggr)^{2\varepsilon} D(\varepsilon)\biggr]^l
+ \mathcal{O}\biggl(\frac{1}{\beta_0^3}\biggr)\biggr\}\,,
\nonumber\\
&&\tilde{F}_\Sigma(\varepsilon,u) = \frac{e^{2 \gamma u}}{\Gamma(1-2\varepsilon-2u)} F_\Sigma(\varepsilon,u)\,.
\label{Ft2}
\end{eqnarray}
The renormalized result does not depend on $\mu$:
\begin{equation}
R_{\sum m_i^2}(\omega) = - \frac{16}{3} \frac{N_c C_F T_F}{(4\pi)^{d/2}} \biggl[
\frac{1}{\beta_0^2} \int_0^\infty du\,e^{-u/b} \tilde{S}_\Sigma(u) + \mathcal{O}\biggl(\frac{1}{\beta_0^3}\biggr)\biggr]\,,\quad
\tilde{S}_\Sigma(u) = \frac{\tilde{F}_\Sigma(0,u)}{u}\,.
\label{RS}
\end{equation}
The function $\tilde {S}_\Sigma(u)$ has no UV renormalon pole at $u=\frac{1}{2}$;
IR renormalon poles are situated at integer $u\ge2$:
\begin{equation}
\tilde{S}_\Sigma(u) \sim - \frac{2}{2-u}\,.
\end{equation}

\section{Conclusion}
We have calculated the coefficient functions up to dimension 2 of the correlator of the heavy-light quark current
in HQET at the leading order in $1/\beta_0$, details of calculations can be found in~\cite{Grozin:2024dhk}.
The UV renormalon ambiguity ($u=\frac{1}{2}$) of the coefficient functions
is compensated by the IR renormalon ambiguity ($u=\frac{1}{2}$) of the pole mass $M$.
IR renormalon ambiguities of the coefficient functions
are compensated by UV renormalon ambiguities of vacuum condensates.

The work was supported by the Ministry of Science and Higher Education of Russia.

\end{document}